\documentclass[fleqn,10pt]{wlscirep}
\usepackage[utf8]{inputenc}
\usepackage[T1]{fontenc}
\usepackage{color}
\usepackage[normalem]{ulem}
\usepackage{subcaption}
\usepackage{url}

\title{On-Demand Virtual Research Environments using Microservices}

\author[1,2,*]{Marco Capuccini}
\author[4]{Anders Larsson}
\author[2]{Matteo Carone}
\author[2]{Jon Ander Novella}
\author[3]{Noureddin Sadawi}
\author[3]{Jianliang Gao}
\author[1]{Salman Toor}
\author[2]{Ola Spjuth}

\affil[1]{Department of Information Technology, Uppsala University, Sweden}
\affil[2]{Department of Pharmaceutical Biosciences, Uppsala University, Sweden}
\affil[3]{Department of Surgery and Cancer, Imperial College London, London, UK}
\affil[4]{National Bioinformatics Infrastructure Sweden, Uppsala University, Sweden}

\affil[*]{marco.capuccini@it.uu.se}

\begin{abstract}
The computational demands for scientific applications are continuously increasing. The emergence of cloud computing has enabled on-demand resource allocation. However, relying solely on infrastructure as a service does not achieve the degree of flexibility required by the scientific community. Here we present a microservice-oriented methodology, where scientific applications run in a distributed orchestration platform as software containers, referred to as on-demand, virtual research environments. The methodology is vendor agnostic and we provide an open source implementation that supports the major cloud providers, offering scalable management of scientific pipelines. We demonstrate applicability and scalability of our methodology in life science applications, but the methodology is general and can be applied to other scientific domains.
\end{abstract}
\begin{document}

\flushbottom
\maketitle
%
%
\thispagestyle{empty}


\section{Introduction}
Modern science is increasingly driven by compute and data intensive processing. Datasets are increasing in size and are not seldom in the range of gigabytes, terabytes or even petabytes, and at the same time large-scale computations may require thousands of cores~\cite{Dubitzky2011}. Hence, access to adequate e-infrastructure represents a major challenge in science. The need for computing power can vary a lot during the course of a research project, and large resources are generally needed only when large-scale computations are being executed~\cite{lampa2013lessons, dahlo2018tracking}. To this extent, moving analyses to cloud resources represents an interesting opportunity. In fact, cloud resources come as a configurable virtual infrastructure that can be allocated and released as needed, with a pay-per-use pricing model \cite{Armbrust2009}. Nevertheless, this way of procuring infrastructure introduces a layer of complexity that researchers may find hard to cope with. Configuring virtual resources requires substantial technical skills~\cite{weerasiri2017taxonomy}, and it is generally a tedious and repetitive task when infrastructure is allocated on demand. Therefore, when running scientific applications on cloud there is a need for a methodology to aid this process. To promote sustainability, this methodology should be generally applicable over multiple research domains, thus allowing to compose working environments from established scientific software components. 

The on demand instantiation of scientific working environments on a ``global virtual infrastructure" was envisioned by Candela et al.~\cite{Candela2013}. These working environments, which comprehensively serve the needs of a community of practice, are referred to as Virtual Research Environments (VREs). Candela et al. envisioned VREs to maximize the reuse of existing components, thus being assembled dynamically from a variety of scientific packages, and they pointed out the importance of automation, resilience and monitoring in these systems. Roth et al.~\cite{roth2011} and Assante et al.~\cite{assante2019gcube} introduced two similar cloud-based implementations of such vision. Both implementations enable to dynamically compose VREs from a collection of scientific applications, which are installed directly on Virtual Machines (VMs). Nevertheless, these approaches have a major limitation. By installing scientific software on VMs without an appropriate isolation mechanism, one will almost inevitably encounter conflicting dependencies~\cite{Williams2016}. In fact, scientific applications often come with a complex environment, where the versions of the dependencies can considerably affect the results of an analysis~\cite{DiTommaso2017nextflow}. 
Hence, VREs need to provide a wide collection of software environments allowing the scientists to select trusted tools and dependencies according to the use case at hand.
Under these settings, conflicts can occur between distinct environments as well as between flavors of the same environment. In addition, it is generally hard to package and distribute such a complex environment in a way that will guarantee seamless instantiation.

The technology that have been recently introduced under the umbrella of microservice-oriented architecture (see Section \ref{sec:background}) is increasingly gaining momentum in science, as it provides an improved mechanism for encapsulating and distributing complete software environments~\cite{Williams2016}. The resulting software components are lightweight, easy and fast to instantiate, and they are isolated by design. Noticeable efforts in leveraging this technology for VREs were made by the PhenoMeNal project (in medical metabolomics)~\cite{peters2018phenomenal}, by the EXTraS project (in astrophysics)~\cite{Dagostino2017}, and by the Square Kilometer Array (SKA) project (in radio astronomy)~\cite{ska}. Based on information from these three research initiatives, here we introduce a generally applicable methodology for on-demand VREs. We stress the term \emph{on-demand} as, in contrast to current practices, our VREs are short-lived and dynamically instantiated as computing power is needed, by using cloud infrastructure. When compared to the inspiring work by Roth et al.~\cite{roth2011} and by Assante et al.~\cite{assante2019gcube}, our methodology provides an improved way of packaging and delivering VRE components by adopting the microservices-oriented architecture. Furthermore, rather than describing a specific VRE implementation which builds on top of an ad hoc software stack, such as gCube by Assante et al.~\cite{assante2019gcube}, we present a high-level methodology that employs industry-trusted technology. In doing this we tackle challenges introduced by software fragmentation and heterogeneity of cloud resources~\cite{weerasiri2017taxonomy, vaquero2019research, ranjan2017orchestrating}, which to the best of our knowledge were not previously investigated in the context of on-demand VREs.

Based on our methodology, we implemented KubeNow: a comprehensive platform for the instantiation of on-demand VREs. KubeNow is generally applicable and cloud-agnostic, meaning that it supports the major cloud providers (thus avoiding vendor lock-in). When comparing KubeNow with microservice architecture installers provided by the IT industry, it is important to consider that KubeNow is designed around the idea of on-demand, short-lived deployments. To this extent, high availability is not crucial while deployment speed is of great importance. The presented methodology and KubeNow have been adopted by PhenoMeNal to enable the instantiation of on-demand VREs for large-scale medical metabolomics.

In summary, our key contributions are as follows.

\begin{itemize}
\item We introduce a general methodology for on-demand VREs with microservices (Section \ref{sec:methodology}).
\item We provide an open source implementation, named KubeNow, that enables instantiating on-demand VREs on the major cloud providers (Section \ref{sec:implementation}).
\item We demonstrate the applicability and the scalability of our methodology by showing use cases and performance metrics from the PhenoMeNal project (Section \ref{sec:eval_analysis}).
\item We evaluate the scalability of KubeNow in terms of deployment speed and compare it with a broadly adopted microservice architecture installer (Section \ref{sec:deploy_scaling}).
\end{itemize}

\section{Microservice-oriented architecture and related technology}
\label{sec:background}
The microservice architecture is a design pattern where complex service-oriented applications are composed of a set of smaller, minimal and complete services (referred to as microservices) \cite{thones2015}. Microservices are independently deployable and compatible with one another through language-agnostic Application Programming Interfaces (APIs), like building blocks. Hence, these blocks can be used in different combinations, according to the use case at hand. This software design promotes interoperability, isolation and separation of concerns, enabling an improved agile process where developers can autonomously develop, test and deliver services.

Software container engines and container orchestration platforms constitute the cutting-edge enabling technology for microservices. This technology enables the encapsulation of software components such that any compliant runtime can execute them with no additional dependencies on any underlying infrastructure \cite{oci2016}. Such software components are referred to as software containers, application containers, or simply containers. Among the open source projects, Docker emerged as the de-facto standard software container engine \cite{Shimel2016}. Along with Docker, Singularity has also seen considerable adoption by the scientific community as it improves security on high-performance computing systems \cite{Kurtzer2017singularity}. Even though container engines like Docker and Singularity serve similar purposes as hypervisors, they are substantially different in the way they function. When running a VM, an hypervisor holds both a full copy of an Operative System (OS) and a virtual copy of the required hardware, taking up a considerable amount of system resources \cite{Vaughan-Nichols2016}. In contrast, software container engines leverage on kernel namespaces to provide isolation, thus running containers directly on the host system. This makes containers considerably lighter and faster to instantiate, when compared to VMs. Nevertheless, containers have a stronger coupling with the OS, thus if they get compromised an attacker could get complete access to the host system \cite{Manu2016}. Hence, in real-world scenarios a combination of both VMs and containers is probably what most organizations should strive towards. 

In current best practices, application containers are used to package and deliver microservices. These containers are then deployed on cloud-based clusters in a highly-available, resilient and possibly geographically disperse manner \cite{Khan2017}. This is where container orchestration frameworks are important as they provide cluster-wide scheduling, continuous deployment, high availability, fault tolerance, overlay networking, service discovery, monitoring and security assurance. Being based on over a decade of Google's experience on container workloads, Kubernetes is the orchestration platform that has collected the largest open source community \cite{Asay2016}. Other notable open source orchestration platforms include Marathon \cite{marathon}, which is built on top of the Mesos resource manager \cite{hindman2011mesos}, and Swarm which was introduced by Docker \cite{naik2016building}.

\section{On-Demand VREs with Microservices}
\label{sec:methodology}
In this section we introduce the methodology that enables on-demand VREs. The methodology is built around the microservice-oriented architecture, and its companion technology. Here we explain our solution on a high level, thus not in connection to any specific software product or vendor. Later in this paper (Section \ref{sec:implementation}) we also show an implementation of this methology that builds on top of widely adopted open source tools and cloud providers.

\subsection{Architecture}

\begin{figure}[h!]
  \centering
  \includegraphics[width=0.7\linewidth]{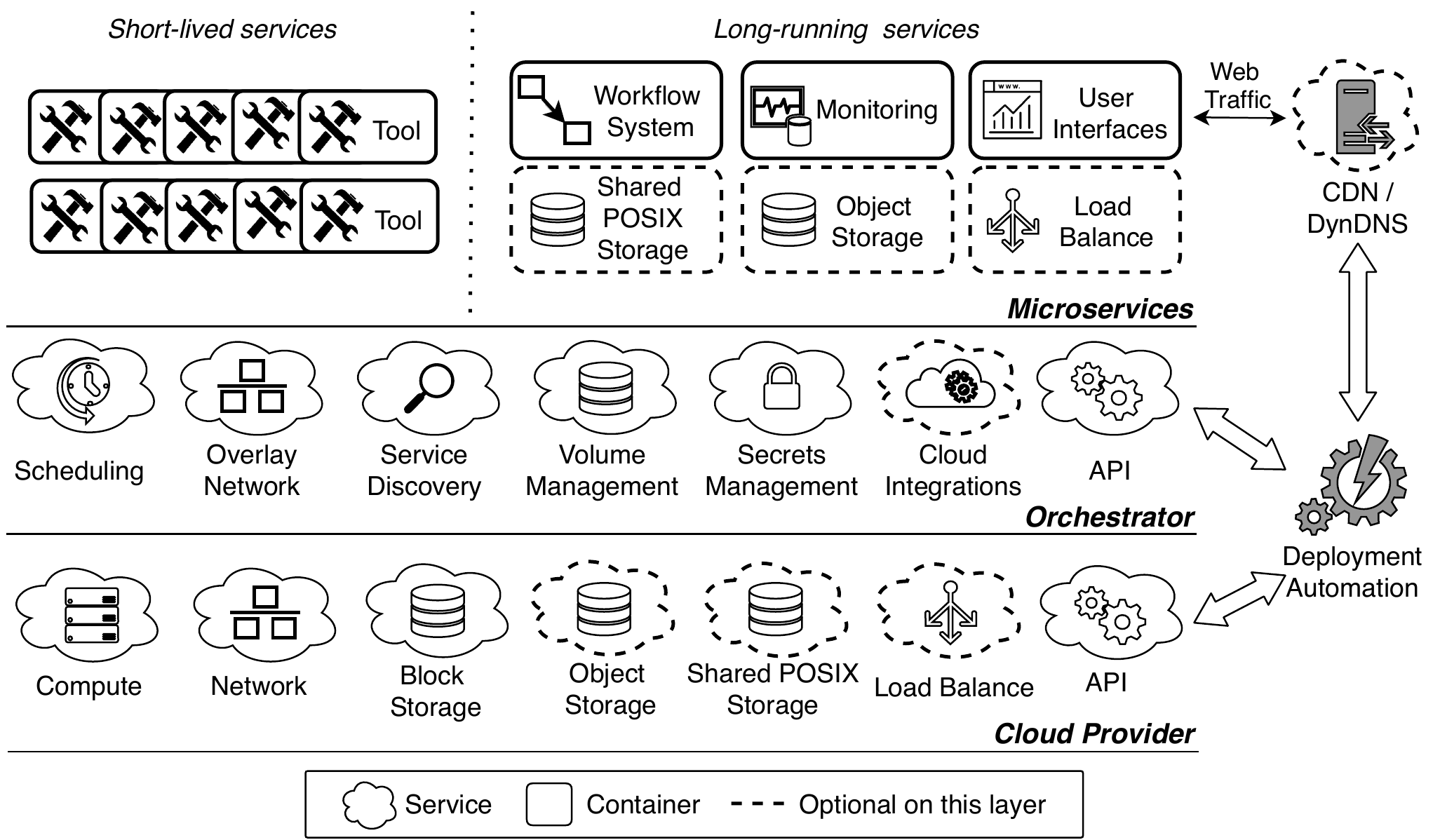}
  \caption{\textbf{Microservice-oriented architecture for on-demand VREs.} The architecture is organized in three layers: \textit{Cloud Provider}, \textit{Orchestrator} and \textit{Microservices}. The two lowest layers offer necessary services to the above layer. In particular the \textit{Cloud Provider} manages virtual resources at infrastructure level, and the \textit{Orchestrator} manages microservices that run as application containers. The uppermost layer run a set of container-based microsrvices for a certain community of practice. The VRE is instantiated through a deployment automation, which may also configure a Content Delivery Network (\textit{CDN}) and a Dynamic Domain Name System (\textit{DynDNS}) to serve the \textit{User Interfaces}.}
  \label{fig:architecture}
\end{figure}

Figure \ref{fig:architecture} shows a general architecture for on-demand VREs. The architecture is organized in three layers: \textit{Cloud Provider}, \textit{Orchestrator} and \textit{Microservices}. In describing each layer we follow a bottom-up approach. 

\subsubsection{Cloud Provider}

At the lowest level, the \textit{Cloud Provider} layer manages virtual resources at infrastructure level. In our methodology this layer enables to dynamically procure infrastructure when a VRE is instantiated. Physical resources can be outsourced (public cloud), in house (private cloud) or anywhere in between (hybrid cloud). 

There are a few necessary services that a cloud system should offer to serve the purpose of a VRE. First, a \textit{Compute} service should enable for booting and managing the VMs that will provide computing power. Second, a \textit{Network} service should provide management for VMs interconnection, routing, security rules and other networking-related concerns. Third, a \textit{Block Storage} service should provide volumes management for VMs. Finally, an \textit{API} should provide programmatic access to the all of the other services (to enable automation). 

Apart from these basic requirements, VREs need a few other services that may not be offered by certain providers (such as moderately sized university installations). Luckily, their implementation as microservices is relatively easy as we describe in Section \ref{sec:uservices}. First, it is important to point out that the main purpose of VREs is to run computations through scientific tools. These tools can be run dispersively in the virtual cluster, thus needing a shared file space for synchronization and concurrent dataset handling. This cannot be provided via block storage, as usually it does not allow for concurrent access. Concurrent access may be achieved via \textit{Object Storage}, a well-established storage service that is capable of providing shared file spaces \cite{karakoyunlu2013}. As the name suggests the service manages files as objects, thus being substantially different from POSIX storage systems. This may represent a challenge in the context of VREs, as scientific tools can usually only operate on a locally-mounted POSIX space. However, this challenge can be tackled by third party products (such as Cloudfuse \cite{cloudfuse}), that can abstract and mount the object storage as a POSIX file system. As an alternative to object storage, some cloud providers recently started to offer \textit{Shared POSIX Storage}, which enables concurrent access on POSIX file spaces. Some examples include Amazon Elastic File System \cite{efs}, Google Cloud File Store \cite{filestore}, Azure NetApp Files \cite{netapp} and OpenStack Manila \cite{manila}. Nevertheless, in contrast to object storage, this solution did not yet reach a consensus in terms of implementation and functionalities across different providers.

Finally, a cloud provider may offer a \textit{Load Balance} service. As the name suggests, this service can be used to load balance incoming traffic from a certain public IP to a configurable set of VMs or microservices. In the context of VREs, this can be useful to expose many services under a single public IP (as related quotas may be limited).

\subsubsection{Orchestrator}
As we mentioned in the introduction, our methodology makes use of application containers to improve the isolation of scientific software environments. When relying solely on cloud providers, VMs represent the most granular mechanism of isolation, and there is no straightforward way to manage disperse containers. This is where the \textit{Orchestrator} is important, as it abstracts VM-based clusters so that containers can be seamlessly scheduled on the underlying resources. There are a few orchestration platforms available in the open source ecosystem (as we discussed in Section \ref{sec:background}), and our methodology is not tied to any of these in particular. However, there are a few services that an \textit{Orchestrator} should offer to support on-demand VREs. 

First, a \textit{Scheduling} service should support cluster-wide resource management and scheduling for application containers. This service should also manage container replication across the cluster, and reschedule failed container (possibly to different nodes in case of VM failure). Since containers can be scheduled across many VMs, an \textit{Overlay Network} should provide interconnection among them. In addition, a \textit{Service Discovery} mechanism should provide the means to retrieve container addresses in the overlay network. This usually comes as a DNS service that should only be available inside the cluster. 

In order to provide data persistency and synchronization between replicas, a \textit{Volume Management} service should offer container volumes operations across the cluster. This means that containers should be able to access a volume, possibly concurrently, on any host. Since this represents a major challenge, on this layer volume management should only represent an abstraction of an underlying storage system, such as a \textit{Block Storage} or a \textit{Shared POSIX Storage}. Apart from file spaces, the \textit{Orchestrator} should be able to manage and mount secrets, such as encryption keys and passwords, in the containers through a \textit{Secret Management} service. 

\textit{Cloud Integrations} may be optionally offered by the orchestrator, and be beneficial in the context of VREs. This service enables to dynamically provision resources on the underlying layer. For instance, on-demand VREs with \textit{Cloud Integrations} may dynamically procure load balancers and cloud volumes for the managed containers. Finally, the \textit{Orchestrator} should provide an \textit{API} to allow programmatic access to its services (enabling automation).

\subsubsection{Microservices}
\label{sec:uservices}
The set of services for a certain community of practice run as container-based microservices, on top of the orchestration platform. While we envision the previous layers to be exchangeable between communities of practice, this layer may offer substantially different functionalities, according to the application domain. Luckily, microservices-oriented systems for different scientific domains (e.g., PhenoMeNal, EXTraS and SKA) are very similar in their design, allowing us to give a general overview of this layer. 

First, we make a distinction between \textit{short-lived services} and \textit{long-running services}. As the name suggests, the former run in the cluster for a limited amount of time while the latter run for the whole life span of the VRE. \textit{Short-lived services} are mainly application containers that run scientific tools, to perform some analyses. The idea consists of instantiating each processing tool, execute a part of the analysis, and allowing it to exit as soon as the computation is done. In this way the analysis can be divided into smaller blocks and distributed over the cluster.

\textit{Long-running services} should include a \textit{Workflow System}, a \textit{Monitoring Platform} and \textit{User Interfaces}. \textit{Workflow Systems} (or similar analytics services) enable to define and orchestrate distributed pipelines of containerized tools. For the containerized tools scheduling to work, it is crucial that the selected workflow system is compatible with the underlying \textit{Orchestrator}. \textit{Monitoring Systems} collect cluster-wide performance metrics, logs and audit trails, possibly aggregating them in visual dashboards. \textit{User Interfaces} provide graphical access to the workflow and monitoring systems, and possibly enable interactive analysis through the execution of live code.

Finally, on this layer \textit{Shared POSIX Storage}, \textit{Object Storage} and \textit{Load Balance} may be implemented as container-based microservices, if not provided by the underlying cloud service. Many available open source projects provide these services and support the major orchestration platforms, thus making the implementation relatively simple (see Section \ref{sec:implementation}). 

\subsubsection{Content Delivery Network and Dynamic Domain Name System}
\label{sec:cdn}

Content Delivery Networks (\textit{CDNs}) are geographically disperse networks of proxy servers \cite{pathan2007taxonomy}. The main goal of a CDN is to improve the quality of web services by caching contents close to the end user. Even though this is not particularly beneficial for short-lived systems, modern CDNs offer additional benefits that are relevant for on-demand VREs. In fact, when proxying web traffic, CDNs can provide seamless HTTPS encryption, along with some protection against common attacks (e.g. distributed denial of service). Since modern CDNs can be configured programmatically via APIs, this provides an easy way to setup encryption on-demand. When comparing with Let's Encrypt \cite{manousis2016shedding}, this system has the advantage of seamlessly issuing and storing a single certificate. This is relevant for on-demand systems, as they may need to be instantiated multiple times in a relatively short period of time, thus making important to reuse existing certificates. In contrast, Let's Encrypt only enables to issue new certificates leaving their management up to the users. 

Dynamic Domain Name System (\textit{DynDNS}) is a method that enables automatic DNS records update \cite{vixie1997dynamic}. Since on-demand VREs are instantiated dynamically, each instance can potentially expose endpoints on different IP addresses. \textit{DynDNS} enables to automatically configure DNS servers, so that endpoints will always be served on a configurable domain name. 

Even though we recommend adoption for user friendliness, \textit{CDNs} and \textit{DynDNS} are optional components. Secure Shell (SSH) tunnelling and virtual private network gateways are valid alternatives to securely access the endpoints. In addition, it is relatively simple to discover dynamically allocated IP addresses by using the cloud API.

\subsubsection{Deployment Automation}
Setting up the presented architecture requires substantial knowledge of the technology, and it may represent a challenge even for a skilled user. Furthermore, for on-demand VREs this time-consuming task needs to be performed for each instantiation. Therefore, on-demand VREs should include a \textit{Deployment Automation}. The automation should operate over multiple layers in the architecture, by setting up the infrastructure through the cloud \textit{API} and by setting up the microservices through the orchestrator \textit{API}. In addition, the automation should also configure the \textit{CDN} and \textit{DynDNS} when required. 

The deployment automation should be based on broadly adopted contextualization tools. These can be cloud-agnostic, thus supporting many cloud providers, or cloud specific. Cloud-agnostic tools are usually open source, while cloud-specific tools may be licensed. The former has the advantage of generalizing operations over many providers, while the latter might offer commercial support. 


\subsection{Continuous Integration}

\begin{figure}[h!]
  \centering
  \begin{subfigure}[b]{0.3\textwidth}
    \includegraphics[width=\textwidth]{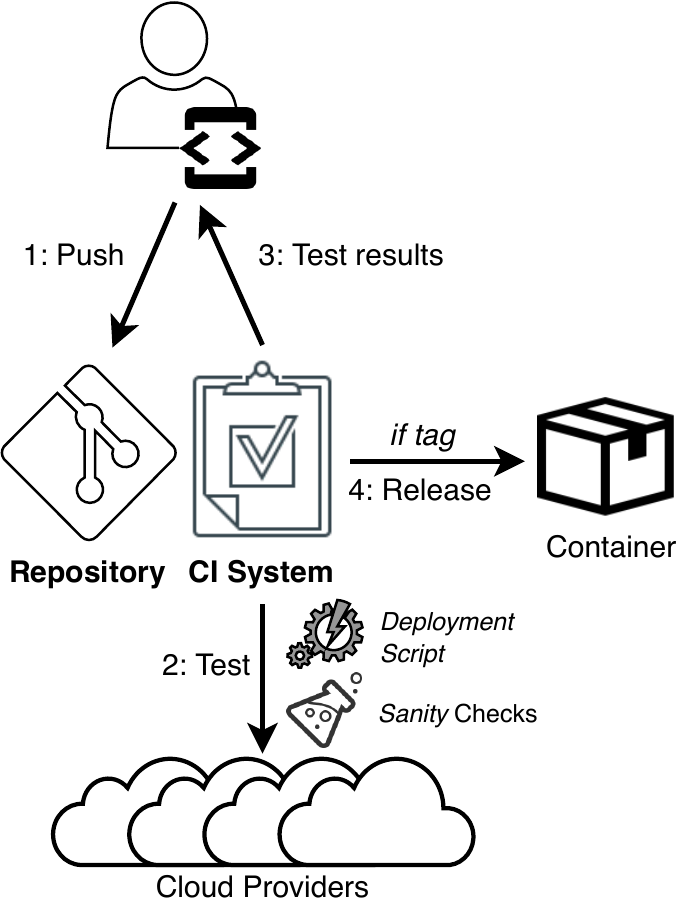}
    \caption{Infrastructure.}
    \label{fig:infra_ci}
  \end{subfigure}
  \hspace{0.15\textwidth}
  \begin{subfigure}[b]{0.3\textwidth}
    \includegraphics[width=\textwidth]{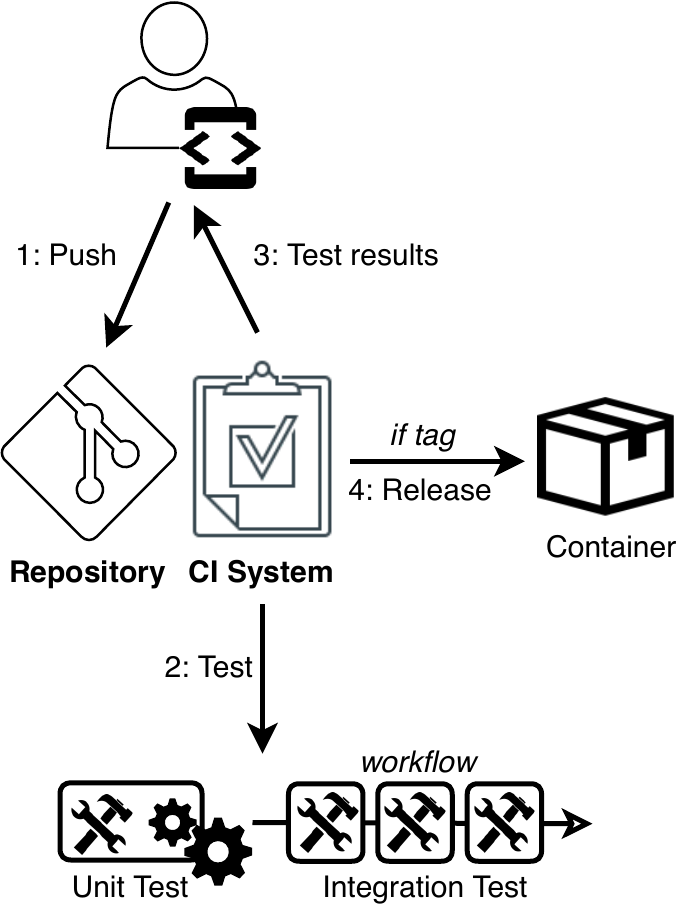}
    \caption{Containerized tools.}
    \label{fig:tools_ci}
  \end{subfigure}
  \caption{\textbf{On-Demand VREs Continuous Integration (CI).} CI should be adopted by communities of practice to aid the collaborative development of virtual infrastructure (figure \ref{fig:infra_ci}) and containerized tools (figure \ref{fig:tools_ci}). Virtual infrastructure is defined through Infrastrucutre as Code (IaC) documents and collaboratively developed using version control. A CI system checks for possible issues on code pushes by running the deployment automation against the supported cloud providers, and thus running sanity checks on the resulting VREs. Similar mechanism is applied to containerized tools, where code pushes trigger unit tests and integration tests on the resulting images. Integration tests consist of testing a workflow that involves the tool. For both infrastructure and containerized tools, if the commit is tagged and the test passes, the code is packaged and released as an application container.}
\end{figure}

Continous Integration (CI) is an agile development practice where software components are integrated frequently through an automation system (also referred to as CI system) \cite{stolberg2009enabling}. The CI system verifies each integration, through unit and integration tests, with the aim of detecting issues as quickly as possible. The communities of practice that develop on-demand VREs should promote collaborative maintenance. To this extent, CI should be adopted for VRE infrastructure (Section \ref{sec:infra_ci}) and containerized tools development (Section \ref{sec:tools_ci}). 

\subsubsection{Infrastructure}
\label{sec:infra_ci}
Infrastructure as Code (IaC) is a methodology that consists in defining virtual infrastructure through machine-readable documents enabling deployment automation, collaborative development and versioning. When developing and maintaining VRE-supporting infrastructure, communities of practice should adopt IaC to enable a robust CI mechanism. 

Figure \ref{fig:infra_ci} shows the CI diagram for on-demand VRE infrastructure. The IaC documents are operated through a version control repository, which is coupled to a CI system. When a collaborator modifies the infrastructure, by pushing a commit to a CI-enabled branch, the CI system automatically runs the deployment automation against the supported cloud providers. After the deployment is completed, the CI system runs some sanity checks on the infrastructure and it returns the results to the collaborator. This happens asynchronously, allowing for the initial push to complete before the testing process. Finally, if the commit is tagged (and the test passes) the CI system packages the virtual infrastructure, along with the dependencies needed for the deployment, as a container image. In this way researchers can easily instantiate the released VRE from their workstation without additional dependencies. 

\subsubsection{Containerized tools}
\label{sec:tools_ci}
Containerized tools are fundamental for on-demand VREs, as they enable to perform scientific analyses. Hence, communities of practice should adopt CI for maintaining container-based distribution of relevant tools.

Figure \ref{fig:tools_ci} shows the CI diagram for a containerized tool. Similarly to infrastructure CI, the container specification is operated through a version control repository, and the testing is triggered by a push to a CI-enabled branch. The testing routine should build the container basing on its specification and run unit tests (as defined by the developers), as well as integration tests. The integration tests consist of running some common workflows that involve the container, possibly detecting introduced issues in these. Finally, if the commit is tagged (and the tests passes) the container is released and made available to the VREs.

\section{Implementation}
\label{sec:implementation}
We provide an open source implementation of our methodology, named KubeNow \cite{kubenow}. KubeNow is generally applicable by design, as it does not explicitly define the uppermost layer in Figure \ref{fig:architecture}. Instead, KubeNow provides a general mechanism to define the microservices layer, so that communities of practice can build on-demand VREs according to their use cases.

KubeNow is cloud-agnostic, and it supports Amazon Web Services (AWS), Google Cloud Platform (GCP) and Microsoft Azure, which are the biggest public cloud providers in the market \cite{Bayramusta2016}, as well as OpenStack (the dominating in-house solution \cite{Elia2017}). This is of great importance in science as it allows to take advantage of pricing options and research grants from different providers, while operating with the same immutable infrastructure. Furthermore, supporting in-house providers enables to process sensitive data, that may not be allowed to leave research centers. 

KubeNow implements \textit{Object Storage}, \textit{Shared POSIX Storage} and \textit{Load Balance} in the microservices layer. This is a straightforward solution to maximize the portability of on-demand VREs. In fact, these services may not be available in certain private cloud installations, and their APIs tend to differ substantially across providers (requiring orchestrators and microservices to be aware of the current host cloud). On the other hand, leveraging on cloud-native services may be beneficial in some cases. As an example, using cloud-native storage enables to persist the data on the cloud, even when the on-demand VRE is not running. Thus, KubeNow gives the possibility to skip the provisioning of \textit{Object Storage}, \textit{Shared POSIX Storage} and \textit{Load Balance}, leaving their handling to the communities of practice in such case.

Finally, KubeNow is built as a thin layer on top of broadly-adopted software products. Below follows a summarizing list.

\begin{itemize}
\item Docker \cite{Shimel2016}: the open source \textit{de facto} standard container engine.
\item Kubernetes \cite{Asay2016}: the orchestration platform that has collected the largest open source community.
\item GlusterFS \cite{gluster}: an open-source distributed file system that provides both shared POSIX file spaces and object storage.
\item Traefik \cite{traefik}: an open-source HTTP reverse proxy and load balancer.
\item Cloudflare\textsuperscript{\textregistered} \cite{cloudflare}: a service that provides CDN and DynDNS.
\item Terraform \cite{terraform}: an open-source IaC tool that enables provisioning at infrastructure level.
\item Ansible \cite{ansible}: an open-source automation tool that enables provisioning of VMs and Kubernetes.
\item Packer \cite{packer}: an open-source packaging tool that enables packaging of immutable VM images.
\item Travis CI \cite{travis}: a CI system that enables KubeNow CI as described in Section \ref{sec:infra_ci}.
\end{itemize}

\subsection{Infrastructure design}

\begin{figure}[h!]
  \begin{center}
  \includegraphics[width=0.6\linewidth]{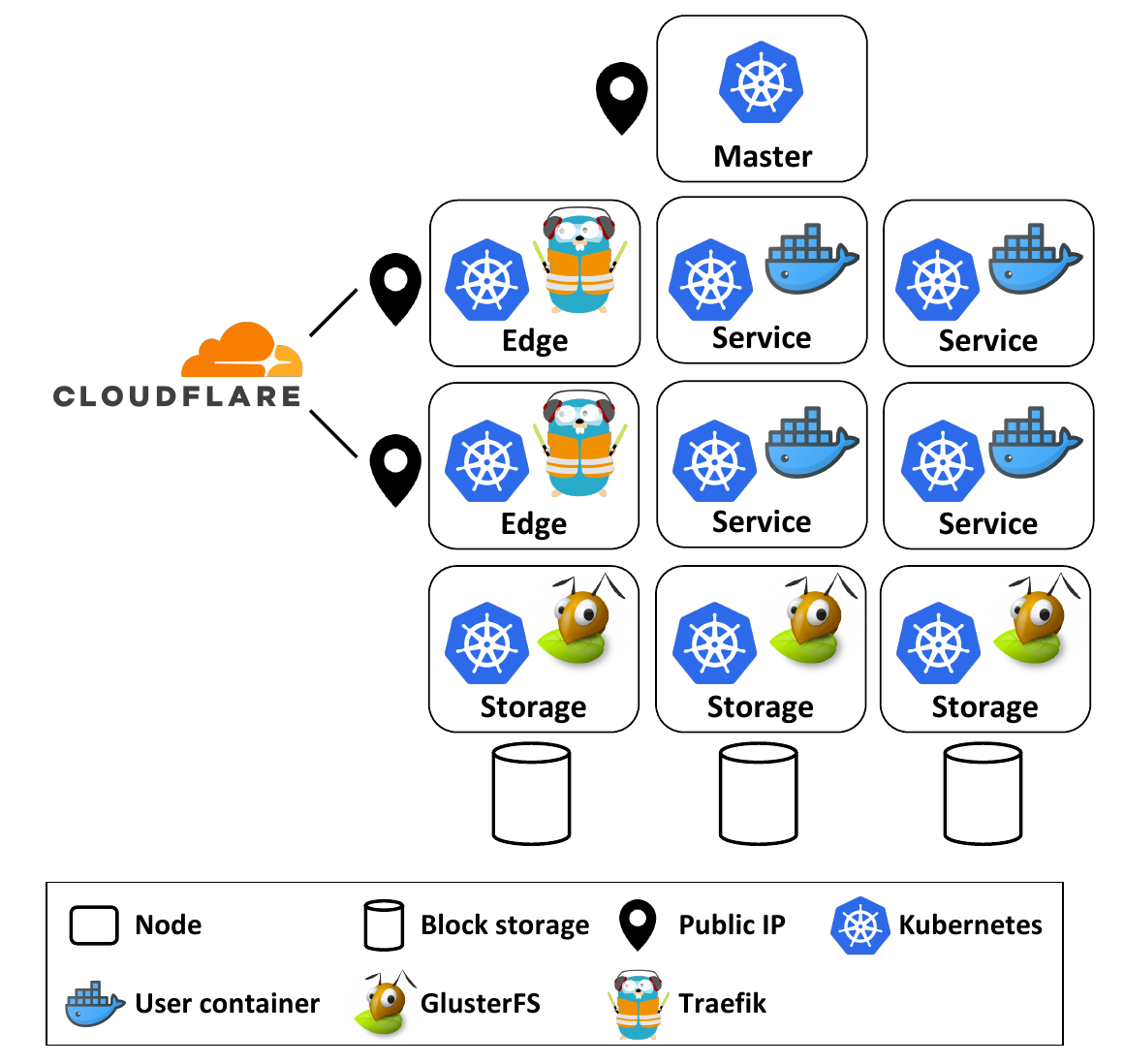}
  \caption{\textbf{KubeNow sample deployment.} There are four main node entities in a KubeNow cluster, which run microservices through Kubernetes. The master node manages various aspects of the other nodes, and it runs the Kubernetes API. Service nodes run the user application containers. Storage nodes run GlusterFS, and they attach a block storage volume to provide more capacity. Edge nodes run Traefik to load balance Internet traffic to the application containers, and each of them is associated to a public IP. Finally, Cloudflare manages DNS records for the edge nodes IP, and optionally proxies Internet traffic to provide encryption.}
  \label{fig:kn_infra}
  \end{center}
\end{figure}

Figure \ref{fig:kn_infra} shows a sample KubeNow deployment at infrastructure level. In a KubeNow cluster there are four main node entities: \textit{master}, \textit{service}, \textit{storage} and \textit{edge}. By default, each node shares the same private network that allows incoming traffic only on SSH, HTTP and HTTPS ports. The master node manages various aspects of the other nodes, retaining the cluster status and running the Kubernetes API. The current implementation of KubeNow does not support multiple master nodes. The purpose of KubeNow is to enable on-demand processing on cloud resources. Under this assumption, deployments are supposed to be short lived, hence high availability is not crucial. Service nodes are general-purpose servers that typically run user containers. Storage nodes run GlusterFS, and they are attached to a block storage volume to provide additional capacity. Finally, edge nodes are service nodes with an associated public IP address, and they act as reverse proxies and load balancers, for the services that are exposed to the Internet. In order to resolve domain names for the exposed services, a wildcard record is configured in the Cloudflare dynamic DNS service \cite{cloudflare}, such that a configurable base domain name will resolve to the edge nodes. In addition, the traffic can be proxied through the Cloudflare servers, using a fully encrypted connection. When operating in this mode Cloudflare provides HTTPS connections to the end user, and it protects against distributed denial of service, customer data breach and malicious bot abuse. 

Apart from the typical setting that we show in Figure \ref{fig:kn_infra}, some other configurations can be used. Excluding the master node, each node entity is optional and it can be set to any replication factor. For instance, when IP addresses are particularly scarce, it is possible to not deploy any edge node, and to use the master node as reverse proxy instead (this may often be the case for private or community cloud settings). The same stands for the storage nodes, that can be removed when an external filesystem is available. In addition, for single-server setups, it is possible to deploy the master node only, and to enable it for service scheduling. Finally, since for entry-level users it can be difficult to reserve a domain name and set it up with Cloudflare, it is possible to use NIP.IO \cite{nip} instead. NIP.IO provides for an easy mechanism to resolve domain names without needing any configuration (e.g., \textit{foo.10.0.0.1.nip.io} maps to \textit{10.0.0.1}, \textit{bar.10.0.0.2.nip.io} maps to \textit{10.0.0.2}, etc.).

\subsection{Deployment automation}
\label{sec:kn_deploy}

Enabling fast and scalable deployments is crucial when leveraging cloud infrastructure on-demand. In fact, if the deployment time grows considerably when increasing the number of nodes, the VRE instantiation time likely dominates over the analysis time, making less appealing to invest in large-scale resources.  

In order to achieve fast and scalable deployments, there are two main ideas that we introduced in our automation. First, the instances are booted from a preprovisioned image (collaboratively developed via Travis CI \cite{travis}). When the image is not present in the cloud user space, the deployment automation imports it, making all of the consecutive deployments considerably faster. Using this approach, all of the required dependencies are already installed in the instances at boot time, without paying for any time-consuming download. The second idea consists in pushing the virtual machines contextualization through cloud-init \cite{cloud-init}, by including a custom script in the instances bootstrap. In this way, the machines configure themselves independently at boot time leading to a better deployment time scaling, when compared to systems where a single workstation coordinates the whole setup process (as we show in Section \ref{sec:evaluation}). This latter approach is even more inefficient when the deployment automation runs outside of the cloud network, which is a quite common scenario.

\begin{figure}[t]
\centering
\begin{verbatim}
    $ kn init <provider> <directory>
    $ cd <directory> 
    $ kn apply
    $ kn helm install <package>
\end{verbatim}
\caption{\textbf{KubeNow CLI user interaction.} The \texttt{init} subcommand sets up a deployment directory for a certain cloud provider, where the user locates and edits some configuration templates. Then, the \texttt{apply} subcommand prepares the KubeNow architecture, and the \texttt{helm} subcommand installs the application-specific research environment.}
\label{fig:kncli}
\end{figure}

The KubeNow deployment automation is available as a Command-Line Interface (CLI), namely \texttt{kn}, that has the goal of making cloud operations transparent. In fact, we envision researchers to autonomously set up cloud resources, without performing complex tasks outside their area of expertise. The \texttt{kn} CLI wraps around a Docker image that encapsulate Terraform, Ansible and a few other dependencies, hence Docker is the only client-side requirement. Figure \ref{fig:kncli} shows a typical user interaction. The user starts by initializing a deployment directory for a certain cloud provider with the \texttt{kn init} command. The deployment directory contains some template files that need to be filled in, to specify a few parameters (e.g., cluster size and credentials). Once the user is done with the configurations, the deployment is started by changing into the deployment directory and by running the \texttt{kn apply} command. This command sets up Kubernetes as well as the KubeNow infrastrucuture (Figure \ref{fig:kn_infra}). Finally, the application-specific research environment is installed on top of KubeNow, by running Helm \cite{helm} (the Kubernetes package manager). Even if preparing Kubernetes packages requires substantial expertise, ready-to-use applications can be made available through Helm repositories. 

\section{Evaluation}
\label{sec:evaluation}
We evaluate our methodology using KubeNow as implementation. Being based on Kubernetes, our system benefits from the resilience characteristics provided by the orchestration platform. Resilience in Kubernetes was previously discussed and studied \cite{vayghan2018deploying, netto2017state, javed2018cefiot} and it is trusted by several organizations \cite{fordThe2016}; thus, we do not show a resiliency evaluation here. We instead show how the adoption of our methodology enable scientific analysis at scale (Section \ref{sec:eval_analysis}) and how KubeNow scales in terms of deployment speed on each cloud provider, also in comparison with a broadly adopted Kubernetes installer (Section \ref{sec:deploy_scaling}). Regarding this last point, it is not our intention to compare the cloud providers in terms of speed or functionalities, but to show that the deployment scales well on each of them.

\subsection{Execution of scientific analysis}
\label{sec:eval_analysis}

KubeNow has been adopted by the PhenoMeNal project to enable the instantiation of on-demand VREs \cite{peters2018phenomenal}. The PhenoMeNal project aims at facilitating large-scale computing for metabolomics, a research field focusing on studying the chemical processes involving metabolites, which constitute the end products of processes that take place in biological cells. On top of KubeNow, the PhenoMeNal VREs run a variety of containerized processing tools as short-lived services as well as three workflow systems, a monitoring platform and various user interfaces. More in detail, the VREs provide Luigi \cite{luigi}, Galaxy \cite{Goecks2010} and Pachyderm \cite{pachyderm} as workflow systems and the Elasticsearch Fluentd Kibana stack \cite{cyvoct2018how} as monitoring platform, all of which come with their built-in user interfaces. In addition, PhenoMeNal VREs also provide Jupyter \cite{Jupyter} to enable interactive analysis through a web-based interface. 

PhenoMeNal VREs have seen applications in mass spectrometry, nuclear magnetic resonance analyses as well as in fluxomics \cite{Emami2017}. Even though these three use cases come from metabolomics studies, they are substantially different and require  different tools and pipelining techniques. This suggests that our methodology is generally applicable and supports applications in other research fields.

\subsubsection{Parallelization of individual tools}
\label{sec:speedup}

Gao et al. \cite{gao2019metabolomics} and Novella et al. \cite{novella2018container} used the PhenoMeNal VREs to parallelize three individual metabolomics tools: Batman \cite{Hao2012}, FeatureFinderMetabo \cite{FeatureFinderMetabo} and CSI:FingerID \cite{duhrkop2015searching}. In these two studies different choices were made in terms of infrastructure setup, utilized workflow system and cloud provider. However, in both cases the parallelization was performed by splitting the data into $N$ partitions, where $N$ was also the number of utilized vCPUs, and by assigning each partition to a containerized tool replica. Gao et al. ran their analysis on 2000 1‐dimensional spectra of blood serum from the MESA consortium \cite{bild2002multi,karaman2016workflow}, while Novella et al. processed a large-scale dataset \cite{herman2018integration} from the Metabolights \cite{haug2012metabolights} repository. 

In both studies the performance is evaluated in terms of measured speedup when increasing the number of utilized vCPUs. The speedup was computed as $T_N/T_1$ where $T_N$ is the running time of the parallel implementation on $N$ cores and $T_1$ is the running time of the containerized tool on single core (measured on the same cloud provider). Gao et al. used the Luigi workflow system to parallelize Batman on Azure and on the EMBL-EBI OpenStack \cite{embassy} installation. When running on Azure they used 10 service nodes with 32 vCPUs and 128GB of RAM each, and 1 storage node with 8 vCPUs and 32GB of RAM. On the EMBL-EBI OpenStack they used 55 worker nodes with 22 vCPUs and 36GB of RAM each, and 5 storage nodes with 8 vCPUs and 16GB of RAM each. Under these settings they run on 50, 60, 100, 250 and 300 vCPUs on Azure, and on 100, 200, 500, 800 and 1000 vCPUs on the EMBL-EBI OpenStack. 

Novella et al. used the Pachyderm workflow system to parallelize FeatureFinderMetabo and CSI:FingerID on AWS. They run their experiments on AWS, using the \textit{t2.2xlarge} instance flavor (8 vCPUs and 32GB of RAM) for each node in their clusters. They used 5 service nodes and 3 storage nodes when running on 20 vCPUs, 8 service nodes and 4 storage nodes when running on 40 vCPUs, 11 service nodes and 6 storage nodes when running on 60 vCPUs, and 14 service nodes and 7 storage nodes when running on 80 vCPUs. 

\begin{figure}[t]
  \begin{center}
  \includegraphics[width=0.7\linewidth]{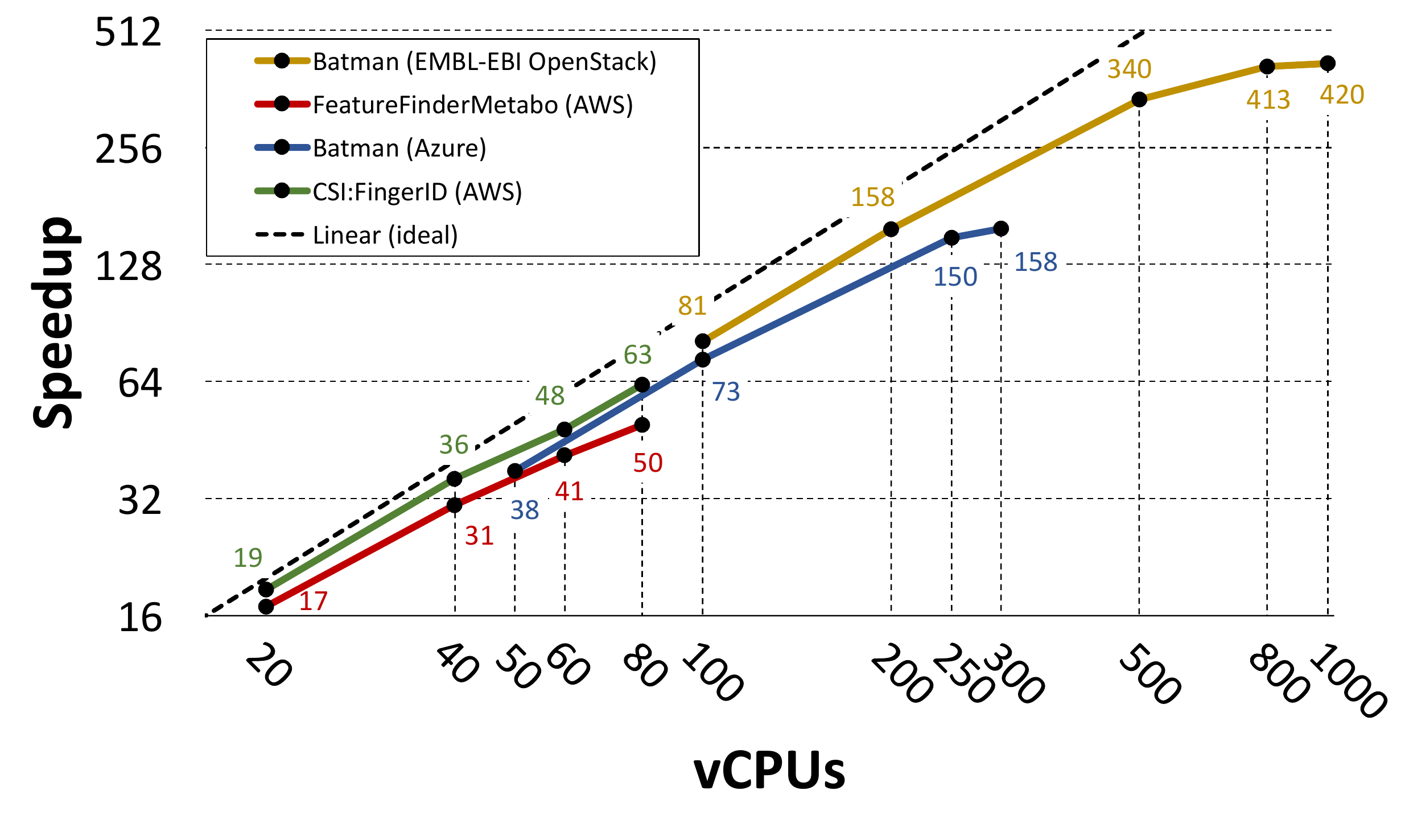}
  \caption{\textbf{Speedup plot for three containerized tools.} The plot shows speedups for three containerized tools that were parallelized using the PhenoMeNal on-demand VRE on different cloud providers. Please notice the logarithmic scale (in base 2) on both axes.}
  \label{fig:speedups}
  \end{center}
\end{figure}

Figure \ref{fig:speedups} shows the measured speedup for each tool in the referenced studies. Even though these tools differ in terms of CPU and I/O demands, their speedup has a close to linear growth up to 500 vCPUs. For the Batman use case, the speedup starts to level out at 300 vCPUs when running on Azure and at 800 vCPUs when running on the EMBL-EBI OpenStack. However, we point out that Gao et al. used only 1 storage node when running on Azure, meaning that in such case more I/O contention occurred. 

\subsubsection{Full analysis scaling}
\label{sec:wse}

Khoonsari et al. \cite{Emami2017} used  the PhenoMeNal VRE to scale the preprocessing pipeline of MTBLS233, one of the largest metabolomics studies available on the Metabolights repository \cite{haug2012metabolights}. This is substantially different from the previous benchmarks, as the analysis was composed by several tools chained into a single pipeline, and because the scalability was evaluated over the full workflow. However, the parallelization was again implemented by assigning a roughly equal split of the data to each container replica. The scalability of the pipeline was evaluated by computing the Weak Scaling Efficiency (WSE) when increasing the number of utilized vCPUs.

\begin{figure}[t]
  \begin{center}
  \includegraphics[width=0.7\linewidth]{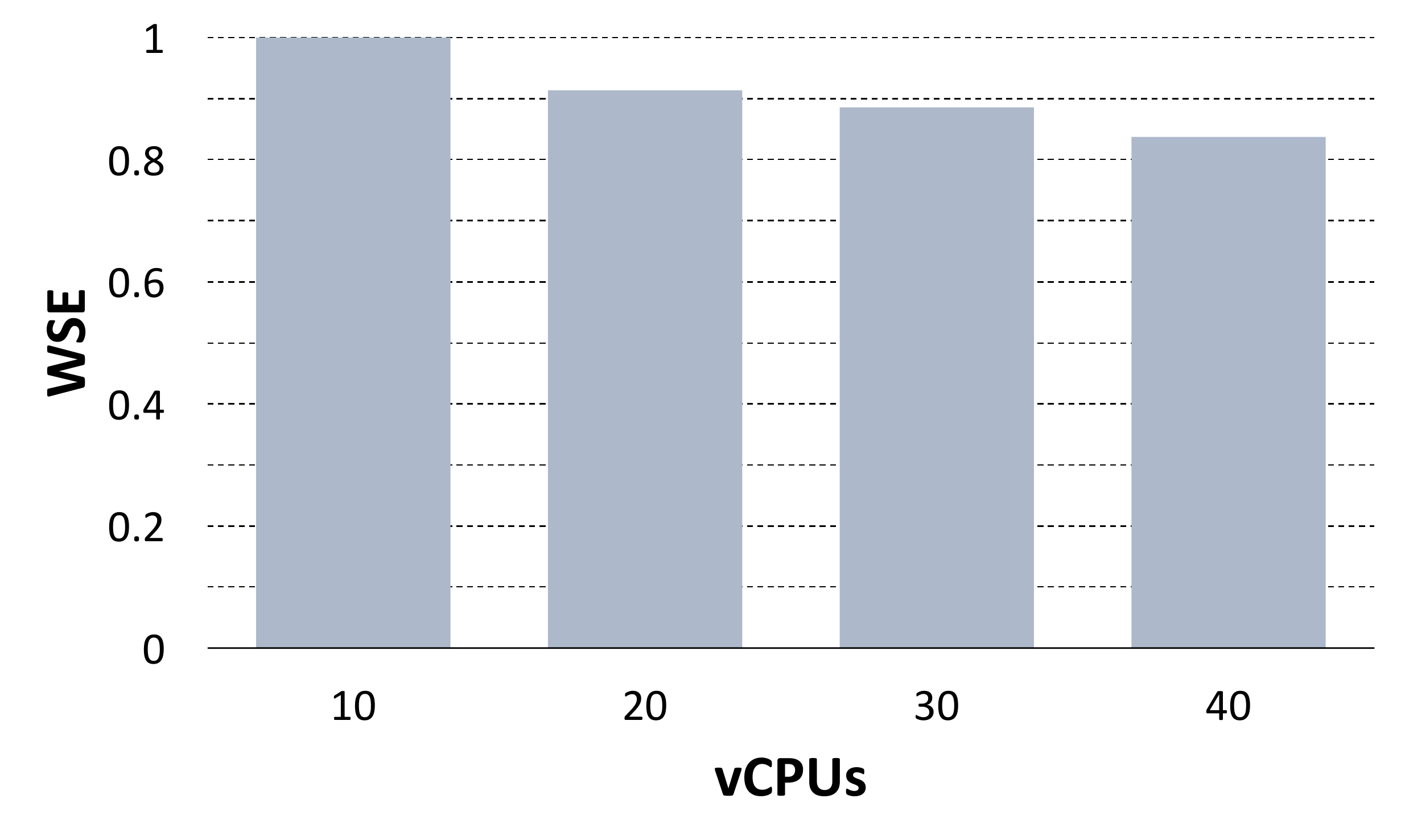}
  \caption{\textbf{WSE plot for the MTBLS233 pipeline.} The plot shows the Weak Scaling Efficiency (WSE) for the MTBLS233 pipeline, executed using the PhenoMeNal on-demand VRE on an OpenStack-based provider.}
  \label{fig:wse}
  \end{center}
\end{figure}

The pipeline was implemented using the Luigi workflow system on the SNIC Science Cloud (SSC) \cite{snic-cloud}, an OpenStack-based provider, using the same instance flavor with 8 vCPUs and 16GB of RAM for each node in the cluster. To compute the WSE, the analysis was repeatedly run on 1/4 of the dataset (10 vCPUs), 2/4 of the dataset (20 vCPUs), 3/4 of the dataset (30 vCPUs) and on the full dataset (40 vCPUs). Then, for $N=10,20,30,40$ the WSE was computed as $T_{10}/T_N$ where $T_{10}$ was the measured running time on 10 vCPUs and $T_N$ was the measured running time on $N$ vCPUs. Figure \ref{fig:wse} shows the WSE measures. There was a slight loss in terms of WSE when increasing the vCPUs, however at full regimen the Khoonsari et al. measured a WSE of $0.83$ indicating good scalability. The loss in WSE is due to growing network contention when increasing the dataset size. This problem can be mitigated by implementing locality-aware scheduling for containers \cite{zhao2018locality}, and we leave this as future work.

\subsection{Deployment automation scalability}
\label{sec:deploy_scaling}

In order to evaluate how KubeNow deployment automation scales over different cluster sizes, we measured and analyzed its deployment time for each of the supported cloud providers: AWS (Frankfurt region), Azure (Netherlands region), GCP (Belgium region) and OpenStack (provided by EMBL-EBI \cite{embassy} and located in the United Kingdom). Then, where applicable, we repeated the measurements using Kubespray \cite{kubespray}, a broadly-adopted Kubernetes cloud installer, to make a comparison. The experiments were carried out from a local laptop, thus envisioning the common scenario where a researcher needs to set up a one-off cluster, in a remote cloud project. More specifically, the laptop was an Apple MacBook Pro (model A1706 EMC 3071) running on the Uppsala University network (Sweden). We measured time for multiple instantiations on the supported cloud providers, doubling the size for each cluster instance. Apart from the size, each cluster had the same topology: one master node (configured to act as edge), and a 5-to-3 ratio between service nodes and storage nodes. This service-to-storage ratio was shown to provide good performance, in terms of distributed data processing, in our previous study \cite{Emami2017}. Hence, we started with a cluster setup that included 1 master node, 5 service nodes and 3 storage nodes (8 nodes in total, excluding master) and, by doubling size on each run, we scaled up to 1 master node, 40 service nodes and 24 storage nodes (64 nodes in total, excluding master). For each of these setups we repeated the measurement 5 times, to consider deployment time fluctuations for identical clusters. Finally, the flavors used for the nodes were: \texttt{t2.medium} on AWS, \texttt{Standard\_DS2\_v2} on Microsoft Azure, \texttt{n1-standard-2} on GCP, and \texttt{s1.modest} on EMBL-EBI OpenStack. 

\subsection{Comparison between KubeNow and Kubespray}

\begin{figure}[t]
  \begin{center}
  \includegraphics[width=0.7\linewidth]{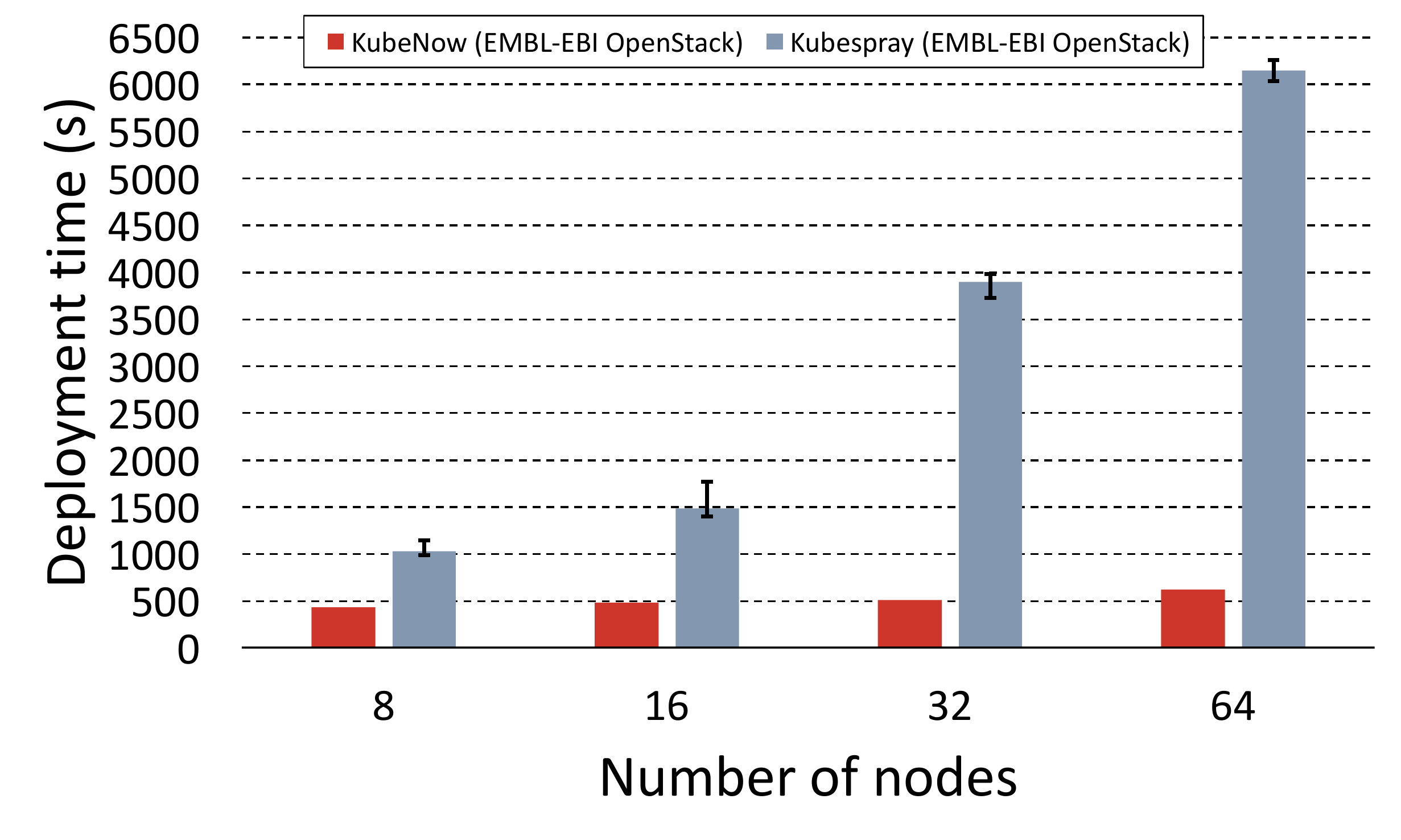}
  \caption{\textbf{KubeNow and Kubespray deployment time comparison.} The plot shows the deployment time, for different cluster sizes (number of nodes), when using KubeNow and when using Kubespray. The experiments were performed on the EMBL-EBI OpenStack. Error bars for KubeNow can be seen on Figure \ref{fig:time_by_cloud}.}
  \label{fig:kn_vs_ks}
  \end{center}
\end{figure}

To make the comparison as fair as possible, we used the Kubespray deployment automation that is based on Ansible and Terraform (the same tools that are used in KubeNow), which uses a bastion node to enable the provisioning with a single IP address. It is worth repeating that public address scarcity is a common issue when dealing with private or community cloud installations, hence we tried to minimize their usage in our experiments. For large deployments, the Kubespray documentation recommends to increase the default maximum parallelism in Ansible and Terraform. Since in our experiments we planned to provision up to 64 nodes, we set the maximum parallelism to this value for both KubeNow and Kubespray. To the best of our knowledge, Kubespray makes storage nodes available only for OpenStack deployments, hence the comparison was possible only on the EMBL-EBI OpenStack provider. Figure \ref{fig:kn_vs_ks} shows the results for KubeNow and Kubespray in comparison.

Deployment time fluctuations for repeated runs, with the same cluster size, were not significant. However, there is a significant difference in terms of scalability between the two systems. In fact, we observe that Kubespray deployments scale poorly, as they increase in time by a large factor when the cluster size doubles. On the other hand, when doubling the number of nodes, KubeNow time increases by a considerably smaller factor, thus providing better scalability. The gap between the two systems becomes of bigger impact as the deployments increase in size. In fact, for the biggest deployment (64 nodes) KubeNow is $\sim$12 times faster than Kubespray. 

To understand why such a big difference occurs, it is important to highlight how the deployment automation differs in the two systems. Kubespray initiates deployments from vanilla images, and it orchestrates the installation from a single Ansible script that runs in the user workstation (outside of the cloud network). Provisioning vanilla images is not only more time consuming, but it also causes more and more machines to pull packages from the same network as the deployments increase in size, impacting scalability. In the same way, the central Ansible provisioner that orchestrates Kubespray's deployments becomes slower and slower in pushing configurations as the number of nodes increases. As we mentioned earlier, KubeNow solves these problems by starting deployments from a preprovisioned image, and by decentralizing the dynamic configuration through cloud-init. 

\subsection{Evaluation on multiple cloud providers}

\begin{figure}[t]
  \begin{center}
  \includegraphics[width=0.7\linewidth]{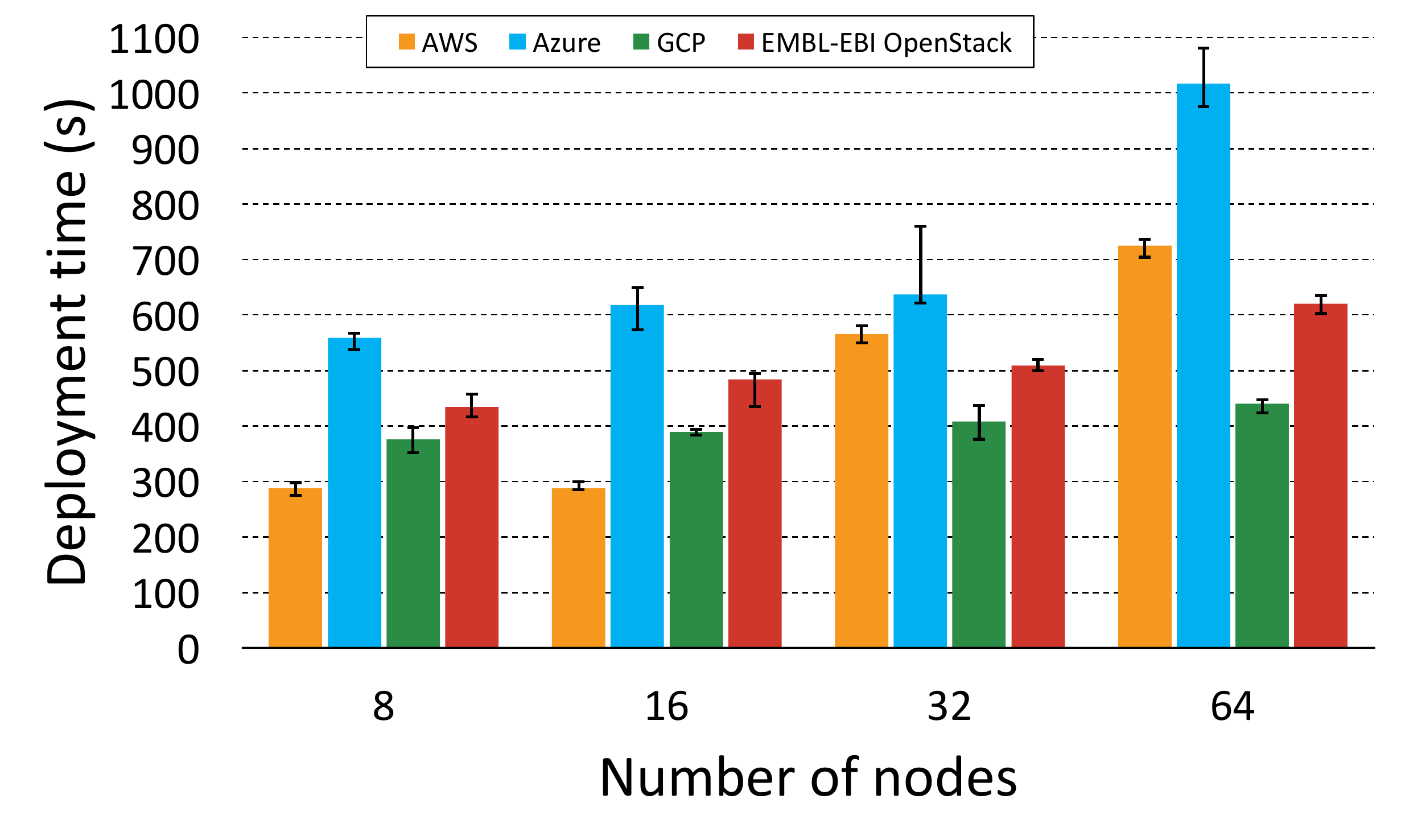}
  \caption{\textbf{KubeNow deployment time by cloud provider.} The plot shows the deployment time for different cluster sizes (number of nodes) on each of the supported cloud providers.}
  \label{fig:time_by_cloud}
  \end{center}
\end{figure}

Figure \ref{fig:time_by_cloud} aims to highlight interesting differences in KubeNow's deployment scaling over different cloud providers. Again, deployment time fluctuations for repeated runs, with the same cluster size, were not significant. We got the best scaling on GCP and EMBL-EBI OpenStack, where every time we doubled the number of provisioned nodes we measured a considerably small increase in deployment time. When deploying on Azure, we always measured a slightly longer time than on the other providers, which increased by a relatively small constant up to 32 nodes. However, when we increased the number of nodes to 64, the deployment time on Azure almost doubled. Finally, on AWS deployment time was better than on the other providers for small clusters (8 and 16 nodes). However, when provisioning 32 and 64 nodes, AWS time increased by a larger factor, and it almost doubled when we scaled from 16 to 32 nodes.

When provisioning on different cloud providers, KubeNow uses the same deployment strategy, which consists in creating the infrastructure with Terraform, and in waiting for the decentralized dynamic configuration to be completed on each node. The same Ansible contextualization is then applied to make small adjustments in the deployment, on every cloud provider. Since the deployment strategy is not cloud-specific, differences in deployment time among clouds are due to the infrastructure layer, which is managed independently by the providers. Finally, it is important to point out that cloud providers can make changes in the infrastructure layer, impacting the results that we show in this study.

\section{Discussion}


The presented methodology differs from the state of the art, as it makes use of the microservice-oriented architecture to deliver on-demand VREs to scientists. This improves isolation of VREs components, and enables to assemble workflows of highly-compartmentalized software components through the adoption of application containers. Achieving scalability by using VMs as isolation mechanism would otherwise be unfeasible, due to the overhead introduced by the guest operating systems.

The implementation for our methodology, namely KubeNow, has been a\-dop\-ted by PhenoMeNal: a live European collaboration in medical metabolomics. Various partners in PhenoMeNal successfully deployed and leveraged KubeNow-based VREs on the major public cloud providers as well as on national-scale OpenStack installations, including those provided by EMBL-EBI \cite{embassy}, de.NBI \cite{denbi}, SNIC \cite{snic-cloud}, CSC \cite{csc} and CityCloud \cite{citycloud}. In addition to KubeNow-based VREs, PhenoMeNal has also implemented the methodology for the CI of containerized tools that we introduced in Section \ref{sec:tools_ci}. Using this methodology PhenoMeNal has continuously delivered $\sim100$ containerized tools for me\-ta\-bo\-lo\-mics processing \cite{peters2018phenomenal}. By referring to use cases in PhenoMeNal, we have shown the ability of our methodology to scale scientific data processing, both in terms of individual tool parallelization (Section \ref{sec:speedup}) and complete analysis scaling (Section \ref{sec:wse}). It is important to point out that since the analyses are fully defined via workflow languages, the pipelines are intrinsically well documented and, by using KubeNow and PhenoMeNal-provided container images, any scientist can reproduce the results on any of the supported cloud providers.

When comparing KubeNow with other available platforms provided by the IT industry, such as Kubespray, it is important to point out that our methodology is conceived for analytics, rather than for highly-available service hosting. This design choice reflects a use case that we envision to become predominant in science. In fact, while the IT industry is embracing application containers to build resilient services at scale, scientists are making use of the technology to run reproducible and standardized analytics. When it comes to long-running service hosting, long deployment time and complex installation procedures are a reasonable price to pay, as they occur only initially. In contrast, we focus on a use case where researchers need to allocate cloud resources as needed. Under these assumptions there is a need for adopting simple, fast and scalable deployment procedures. KubeNow meets these requirements by providing: (1) an uncomplicated user interaction (see Section \ref{sec:kn_deploy}) and (2) fast and scalable deployments (see Section \ref{sec:deploy_scaling}). 


Microservices and application containers are increasingly gaining momentum in scientific applications \cite{peters2018phenomenal, Dagostino2017, ska, Williams2016}. When it comes to on-demand VREs the technology presents some important advantages over current systems. Our methodology is based on publicly available information by three research initiatives in substantially different scientific domains (PhenoMeNal, EXTraS and SKA). It is important to point out that EXTraS and SKA provide microservices-oriented VREs primarly as long running platforms, and they do not cover on-demand instantiation, while our methodology made this possible in PhenoMeNal. The requirements in terms of VRE infrastructure are similar across domains, which allowed us to design our methodology as generally applicable. Hence, we envision our work and the presented benchmarks as valuable guidelines for communities of practice that need to build on-demand VRE systems.

\section{Conclusion}
Here, we introduced a microservice-oriented methodology where scientific  applications  run  in  a  distributed orchestration  platform  as  light-weight  software  containers, referred to as on-demand VREs. Our methodology makes use of application containers to improve isolation of VRE components, and it uses cloud computing to dynamically procure infrastructure. The methodology builds on publicly available information by three research initiatives, and it is generally applicable over multiple research domains. The applicability of the methodology was tested through an open source implementation, showing good scaling for data analysis in metabolomics and in terms of deployment speed. We envision communities of practice to use our work as a guideline and blueprint to build on-demand VREs.

\section*{Data Availability}
The data in the study by Gao et al. \cite{gao2019metabolomics} is publicly available: \url{https://doi.org/10.6084/m9.figshare.c.4204022}. Novella et al. \cite{novella2018container} and Khoonsari et al. \cite{Emami2017} used public data from the Methabolights repository \cite{haug2012metabolights}, and in particular datasets: MTBLS558 and MTBLS233. 

\section*{Ethical approval and informed consent}
Human-derived samples in the datasets are consented for analysis, publication and distribution, and they were processed according to the ELSI guidelines \cite{sariyar2015sharing}. Ethics and consents are extensively explained in the referenced publications \cite{gao2019metabolomics, herman2018integration, ranninger2016improving}.

\bibliography{sample}

\section*{Acknowledgments}
This research was supported by The European Commission's Horizon 2020 programme under grant agreement number 654241 (PhenoMeNal) and the Nordic e-Infrastructure Collaboration (NeIC) via the Glenna2 and Tryggve2 projects. We kindly acknowledge contributions to cloud resources by SNIC, EMBL-EBI, CityCloud, CSC, AWS and Azure. The founders had no role in study design, data collection and analysis, decision to publish, or preparation of the manuscript.

\section*{Author contributions statement}
MCap and OS conceived the project. MCap, AL and MCar implemented the methodology. MCar, JN, NS and JG carried out the evaluation experiments. ST contributed with expertise in cloud computing. All authors read and approved the final manuscript.

\section*{Additional information}
\subsection*{Declaration of interest}
The authors declare that they have no competing interests.

\end{document}